\documentclass[prb,twocolumn,showpacs,preprintnumbers,superscriptaddress,floatfix]{revtex4}

\usepackage[english]{babel}
\usepackage[pdftex]{graphicx}   
\usepackage{dcolumn}            
\usepackage{bm}                 
\usepackage{units}
\usepackage{amsmath}
\usepackage{amssymb}

\newcommand{\kp}{\ensuremath{\boldsymbol{k}_{||}}}

\begin{document}

\title{Tailoring tunnel magnetoresistance by ultrathin\\
       Cr and Co interlayers:\\
       A first-principles investigation of Fe/MgO/Fe junctions}

\author{Peter \surname{Bose}}
\affiliation{Martin Luther University Halle-Wittenberg, Institute of Physics, D-06099 Halle (Saale), Germany}
\affiliation{International Max Planck Research School for Science and Technology of Nanostructures, Weinberg 2, D-06120 Halle (Saale), Germany}

\author{Peter \surname{Zahn}}
\affiliation{Martin Luther University Halle-Wittenberg, Institute of Physics, D-06099 Halle (Saale), Germany}

\author{J\"urgen \surname{Henk}}
\affiliation{Max Planck Institute of Microstructure Physics, Weinberg 2, D-06120 Halle (Saale), Germany}

\author{Ingrid \surname{Mertig}}
\affiliation{Martin Luther University Halle-Wittenberg, Institute of Physics, D-06099 Halle (Saale), Germany}
\affiliation{Max Planck Institute of Microstructure Physics, Weinberg 2, D-06120 Halle (Saale), Germany}
  
\date{\today}

\begin{abstract}
  We report on systematic ab-initio investigations of Co and
  Cr interlayers embedded in Fe(001)/MgO/Fe(001) magnetic tunnel junctions,
  focusing on the changes of the electronic structure and the
  transport properties with interlayer thickness. The results of
  spin-dependent ballistic transport calculations reveal options to
  specifically manipulate the tunnel magnetoresistance ratio.  
  The resistance area products and the tunnel magnetoresistance ratios
  show a monotonous trend with distinct oscillations as a function of the Cr thickness. 
  These modulations are directly addressed and interpreted by means of 
  magnetic structures in the Cr films and by complex band structure 
  effects. The characteristics for embedded Co
  interlayers are considerably influenced by interface resonances
  which are analyzed by the local electronic structure.
\end{abstract}


\maketitle

\section{Introduction}
\label{sec:intro}
During the last years magnetoresistive effects --- in particular the
tunnel magnetoresistance (TMR) effect\cite{Julliere1975,Moodera1995}
--- became increasingly important for the fast developing field of
spintronic devices\cite{Zutic2004,Wolf2006}.  The first industrial
applicable TMR contacts have been built using crystalline MgO
insulators which are epitaxially grown on as well as coated with iron
electrodes\cite{Gallagher2006,Yuasa2007}. Fe/MgO/Fe magnetic tunnel
junctions (MTJs) have been extensively investigated to elucidate the
mismatch between theoretically predicted\cite{Butler2001,Mathon2001}
and the at least one order of magnitude smaller
measured\cite{Yuasa2004,Parkin2004} TMR ratios.  It turned out that
the disparity can be attributed to differences between idealized (in
theory) and real (in experiment) samples.  More sophisticated theories
which include imperfections, like interface
disorder\cite{Zhang2003a,Mathon2006,Itoh2006,Waldron2006,Heiliger2007a,Bose2008,Bonell2009}
or roughness effects\cite{Xu2006}, were able to close the gap between
experiment and theory and highlight the importance of perfect
interfaces.

Although other tunnel junctions, like CoFeB/MgO/CoFeB MTJs with their
high TMR ratios\cite{Ikeda2008}, were put into the focus of attention,
Fe/MgO/Fe MTJs are still intensely studied. Besides the emerging field
of spin-torque effects\cite{Matsumoto2009a}, research is ongoing in
search of other ways to increase the TMR ratio further.  Instead of
improving the interface quality an alternative means is found in the
specific manipulation of the spin-dependent conductances by embedding
ultrathin interlayers\cite{Belashchenko2005,Greullet2007}.

The insertion of a single layer-wise antiferromagnetic (LAFM) Cr
interlayer results into even-odd oscillations of the TMR ratio as a
function of the Cr thickness\cite{Nagahama2005, Matsumoto2009}. In
this paper we report on a first-principles study of these transport
characteristics. We discuss the origin of these
modulations with the apparent $\unit[2]{ML}$-wavelength as well.

Additionally, an analysis of the electronic transport results for Co
interlayers at both Fe/MgO interfaces is presented. These
investigations were motivated by previous ab-initio
calculations\cite{Zhang2004a} which predicted larger TMR ratios for
MgO tunnel junctions with bcc Co(001) leads instead of Fe(001)
electrodes. Due to the fact that Co grows epitaxially only up to a few
monolayers on bcc substrates, a question arises whether ultrathin Co
interlayers could be alternatively used to obtain an enhancement of
the TMR ratios in Fe/MgO/Fe MTJs. To answer this question we computed
the conductances and TMR ratios for small Co interlayer thicknesses
and analyzed the results by means of the electronic structures.

\section{Theoretical Background}
\label{sec:theory}
Our computational approach is divided into two steps. Firstly, the
electronic structures of the MTJs are calculated from first
principles. Secondly, the electronic transport properties are
computed, using the potentials obtained in the first step.
 
The electronic structure is determined self-consistently within the
framework of density-functional theory (DFT) using a
scalar-relativistic screened Korringa-Kohn-Rostoker (KKR) Green
function technique\cite{Zeller1995,Papanikolaou2002}. The spherical
site potentials were treated in the atomic sphere approximation (ASA)
using the local spin density approximation (LSDA) for the
exchange-correlation potential\cite{Kohn1965}. Throughout this work
a parameterization following Vosko, Wilk, and Nusair\cite{Vosko1980}
was used.

Since structural information of Fe/MgO/Fe MTJs with embedded ultrathin
Cr and Co spacers are not reported so far, we resort to a geometry of
planar Fe(001)/MgO/Fe(001) junctions which has been determined
experimentally by surface x-ray diffraction analyses
\cite{Meyerheim2002,Tusche2005}. This structure has been used in previous
theoretical studies\cite{Heiliger2005,Bose2008}. In detail, a
supercell geometry with six MgO layers sandwiched by 20 Fe layers was
used to compute self-consistently the atomic potentials.  The
insertion of $x$ magnetic interlayers in planar
Fe(001)/$x$(Cr)/6(MgO)/Fe(001) and
Fe(001)/$x$(Co)/6(MgO)/$x$(Co)/Fe(001) junctions was achieved by
replacing $x$ Fe monolayers (ML) at the Fe/MgO interfaces by $x$ Cr or
Co layers. This procedure implies that Cr and Co atoms occupy the same
positions as the replaced Fe atoms; worded differently, the
interlayers follow the bcc structure of the Fe(001) leads.

Due to the broken translational invariance in transport direction
($z$, i.\,e.\ [001]) and the in-plane translational invariance, the
eigenstates of the electrodes are labeled by in-plane wave vectors
$\kp = (k_x, k_y)$.  The point group of the two-dimensional lattice is
$4mm$.

The ballistic conductance $C$ per unit cell area $A_{\Box}$ is
computed for zero bias voltage in terms of transmission probabilities
(Landauer-B\"uttiker approach\cite{Imry1999}) at the Fermi energy
$E_F$,
\begin{align}
  C & = \frac{e^2}{h}\int_{\mathrm{2BZ}} T(\kp, E_F) d\kp{}.
  \label{eq:conductance}
\end{align}
The transmission probability $T(\kp, E_F)$ is obtained by means of a
Green function approach\cite{Henk2006}. The integration over the
two-dimensional Brillouin zone (2BZ) requires typically about 90\,000
$\kp$.  The use of special $\kp$ points\cite{Evarestov1983} reduces
that number to $1/8$. The resistance area product
\begin{align}
  RA & = \frac{1}{C},
\end{align}
as normalized quantity, is used to compare the theoretical with
experimental data.
   
The optimistic TMR ratio is obtained from the RAs which are computed
for the parallel ($\mathrm{P}$) and the anti-parallel ($\mathrm{AP}$)
alignment of the two Fe(001) lead magnetizations,
\begin{align}
  \mathrm{TMR} & =
  \frac{RA^{\mathrm{AP}}-RA^{\mathrm{P}}}{RA^{\mathrm{P}}}.
  \label{eq:tmr}
\end{align}
For the normalized TMR ratio the denominator is replaced by 
$RA^{\mathrm{AP}}+RA^{\mathrm{P}}$.
Since interfaces determine considerably the transport properties,
transmittance maps which display $T(\kp, E_F)$ versus $\kp$ need to be
interpreted by means of the local electronic structure, rather than by
the electronic structures of the bulk electrodes.  The former is
obtained from the layer-resolved Bloch spectral density (SD)
\begin{align}
  N_{al}(E,\kp) & = -\frac{1}{\pi} \mathrm{Im}\mathrm{Tr} G^+_{al}(E,\kp),
  \label{eq:sd}
\end{align}
of atom $a$ in layer $l$. $G^+_{al}(E,\kp)$ is the site-diagonal Green
function of that site.  The trace involves integration over the ASA
sphere and summation over spin-angular quantum numbers.

\section{Results}
\label{sec:results}

\subsection{Cr interlayer in Fe(001)/MgO/Fe(001)}
\label{subsec:Cr}
In the following we present results of the thickness dependence of
both the conductances and the TMR ratios for ultra-thin Cr interlayers
which are embedded at a single interface in
Fe(001)/$x$(Cr)/6(MgO)/Fe(001) MTJs. The Cr thickness
$d_{\mathrm{Cr}}$ is varied in steps of monolayers (ML), $x = 0,
\ldots, 7$ with $d_{\mathrm{Cr},\unit[1]{ML}}=\unit[2.35]{\AA}$,
$d_{\mathrm{Cr},\unit[2]{ML}}=\unit[4.04]{\AA}$ and
$d_{\mathrm{Cr},\unit[x]{ML}}=d_{\mathrm{Cr},\unit[2]{ML}}+x\cdot\unit[1.69]{\AA}$.

\begin{figure}
  \includegraphics[width=0.95\columnwidth]{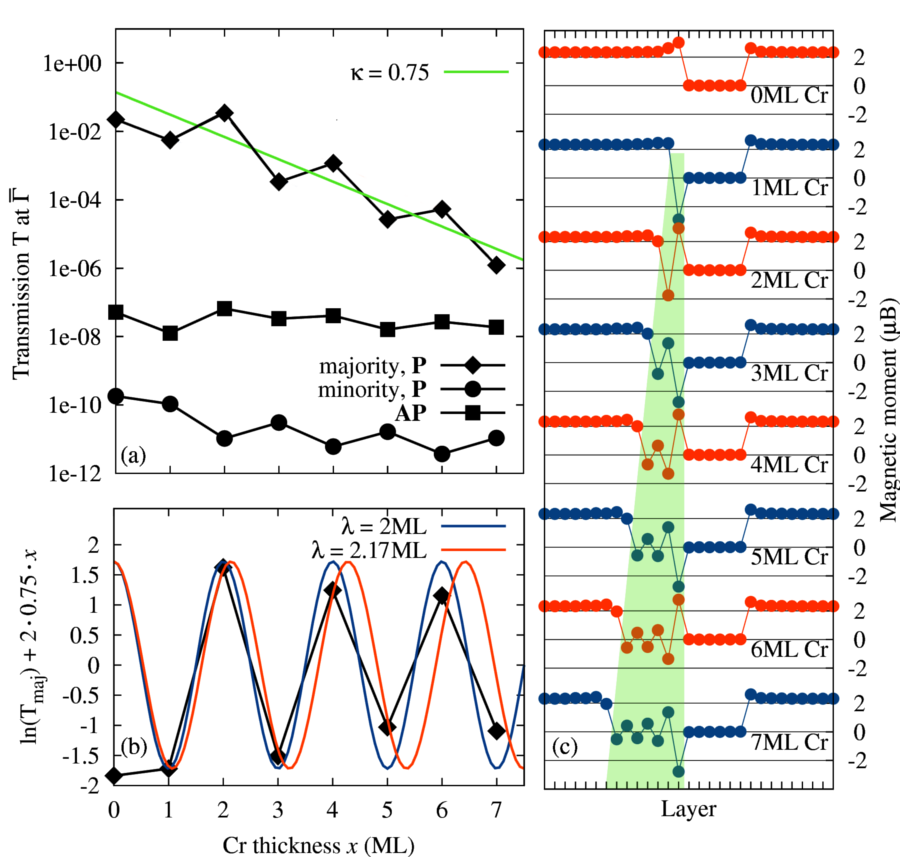}
  \caption{(Color) (a) Dependence of spin-resolved $\mathrm{P}$
    (majority, minority) and $\mathrm{AP}$ transmissions on a
    logarithmic scale versus Cr thickness $x$ for Bloch states at $\kp
    = 0$ in Fe(001)/$x$(Cr)/6(MgO)/Fe(001) MTJs, $x = 0, \ldots, 7$.
    The green line is a fitted exponential to the majority
    transmission. The deviation of the majority transmission from this
    fit is shown in panel (b).  These data are fitted by cosine functions
    (fixed at $x = 1$) with \unit[2]{ML} (blue) and
    \unit[2.17]{ML} (red) periods.  (c) Magnetic profiles of MTJs
    with Cr layer thicknesses $x = 0, \ldots, \unit[7]{ML}$. The green
    area highlights the magnetic moments of the Cr layers.}
    \label{fig:cr_conductance_gamma}
\end{figure}

We start with the tunneling behavior of Bloch states at $\bar{\Gamma}$
($\kp = 0$). The associated transmission probabilities for $x = 0,
\ldots, 7$ are plotted in Fig.~\ref{fig:cr_conductance_gamma}a.  Both
the minority spin contribution for $\mathrm{P}$ and the $\mathrm{AP}$
contribution stay almost constant, whereas the majority spin
contribution of $\mathrm{P}$ decays exponentially with an oscillatory
modulation. Its decay rate is estimated by an exponential fit,
$\exp(-2\kappa x)$ with $\kappa = 0.75$ (green line in
Fig.~\ref{fig:cr_conductance_gamma}a).

It turns out that an oscillatory modulation of
$T_{\mathrm{maj}}^{\mathrm{P}}$ shows up for all $\kp$ within the
2BZ\@. These oscillations are also present in the conductance $C$
which is an integral over the transmission probabilities,
eq.~(\ref{eq:conductance}); hence, there is no destructive
interference which would lead to (complete) cancellation. Thus, it is
essential to elucidate the underlying mechanism of these oscillations.
To strengthen the discussion we focus in the following on the
transmissions at $\bar{\Gamma}$.
 
The oscillation period can be estimated by fitting cosine functions to
the deviation of $T_{\mathrm{maj}}^{\mathrm{P}}(\kp=0)$ from the
averaged exponential decay (black diamonds in
Fig.~\ref{fig:cr_conductance_gamma}b). The fit with a period of
$\unit[2]{ML}$ (blue) reproduces only the peak positions but deviates
significantly in amplitude. A second fit, with $\unit[2.17]{ML}$ period
(red), hits the data best.

\begin{figure}
  \includegraphics[width=0.9\columnwidth]{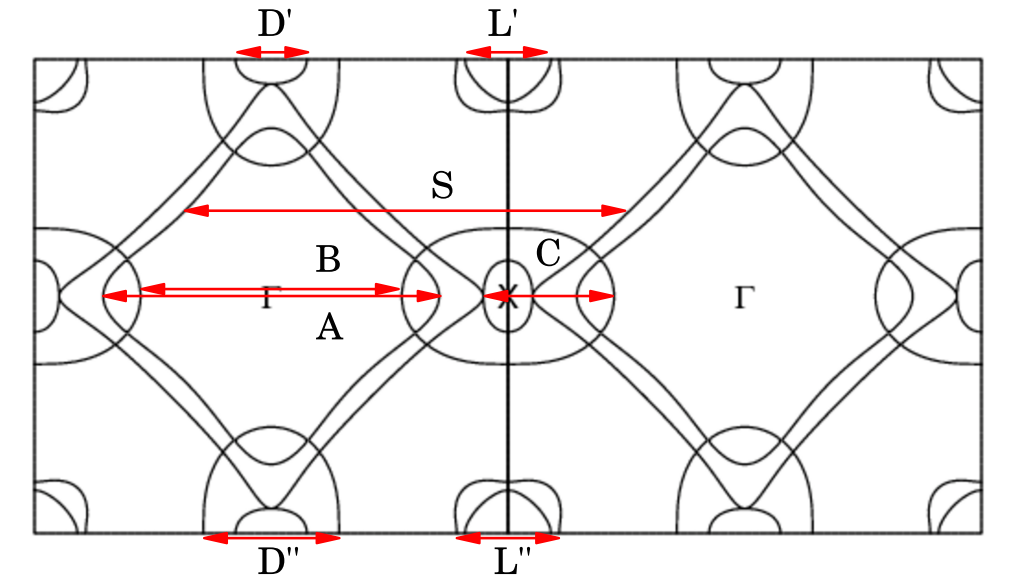}
  \caption{(Color online) Fermi surface cross sections in the (100)
    plane of commensurate AFM Cr (c-afm). The nesting vectors along
    [001] (arrows) are shown in an extended zone scheme and are listed
    in Table~\ref{tab_nesting}.}
  \label{fig:fecrmgofe_fs}
\end{figure}

\begin{table}
  \caption{Nesting vectors of commensurate AFM (c-afm) Cr along [001], as given in Fig.~\ref{fig:fecrmgofe_fs},
    are characterized by their oscillation periods (in $\unit{ML}$).} 
  \begin{tabular}{ccccccccc}
    \hline\hline
      c-afm                 &    L' &  L''  &   S  &   A  &   B  &   C  &    D' & D''  \\ \hline
      $\lambda [\unit{ML}]$ & 11.03 & 9.21 & 2.12 & 2.82 & 3.62 & 7.21 & 13.26 &  6.96 \\
      \hline\hline
  \end{tabular} 
  \label{tab_nesting} 
\end{table}

The $\unit[2]{ML}$ oscillation can be explained by the local magnetic
structure of the layer-wise antiferromagnetic (LAFM) Cr interlayers
(Fig.~\ref{fig:cr_conductance_gamma}c). The magnetic moments of the Cr
layers at the Cr/MgO interface is sizably enhanced due to the
nonmagnetic MgO film.  As a consequence of these partially
uncompensated magnetic moments, the Cr interface layers act as
spin filters for the tunnel currents. Hence, the
latter are increased (decreased) if the magnetic moments within the Cr
interface layers are parallel (anti-parallel) to the
magnetization of the opposite Fe electrode. Due to the LAFM growth of the Cr interlayer
the tunnel current characteristics should exhibit signatures of
$\unit[2]{ML}$ oscillations. The maxima of these oscillations 
should arise for $\mathrm{P}$ ($\mathrm{AP}$) magnetic configurations
of the Fe leads at even (odd) multiples of the Cr interlayer thickness $x$.
This behaviour was already found for LAFM Mn interlayers in Fe(001)/$x$(Mn)/Vac/Fe(001)
MTJs\cite{Bose2007}.  However, the mismatch in
Fig.~\ref{fig:cr_conductance_gamma}b appears like an undersampling
which cannot be satisfactorily explained by means of the
spin filter effect. The better match of the other oscillation period with $\unit[2.17]{ML}$
points to another effect which additionally influences the electronic transport.

\begin{figure}
  \includegraphics[width=0.8\columnwidth]{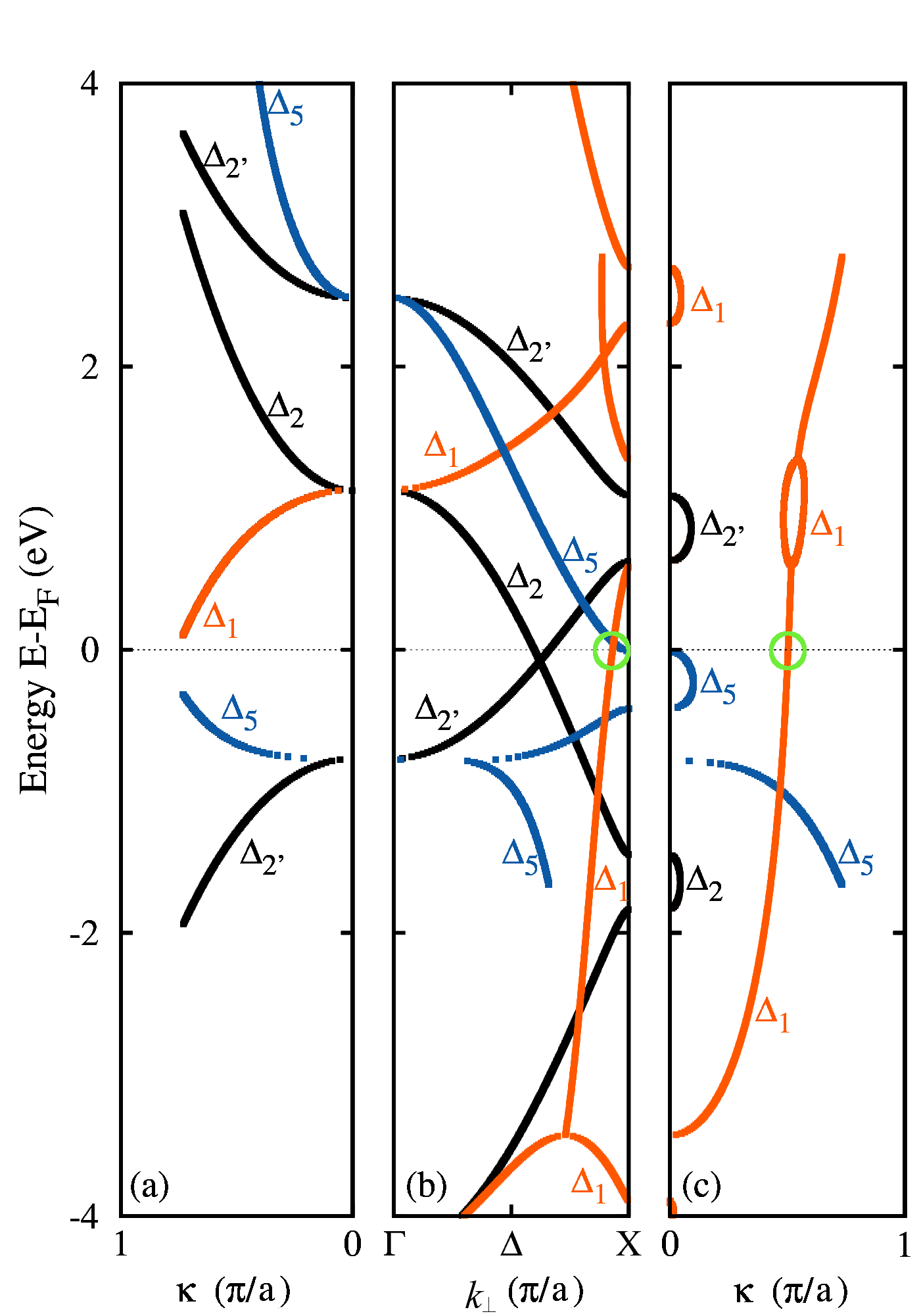}
  \caption{(Color) Complex bandstructure of bulk bcc Cr in the
    commensurate AFM (c-afm) phase along [001] for $\kp = 0$.  The
    real part of $k_{\perp}$ and the exponential decay rate $\kappa$
    (imaginary part of $k_{\perp}$) are shown in panels (b) and (a,
    c), respectively.  The color code of the bands indicates the
    irreducible representations of the point group $4mm$ of the
    associated Bloch states. The green circles at the Fermi energy
    mark the complex $\Delta_1$-band of second kind, to
    $k_{\perp} = (0.922, 0.503)\nicefrac{\pi}{a}$, which governs the
    transmission of majority electrons at $\bar{\Gamma}$ (cf.\
    Fig.~\ref{fig:cr_conductance_gamma}a).}
  \label{fig:fecrmgofe_bs}
\end{figure}

\begin{figure*}[tb]
  \includegraphics[width=0.75\textwidth]{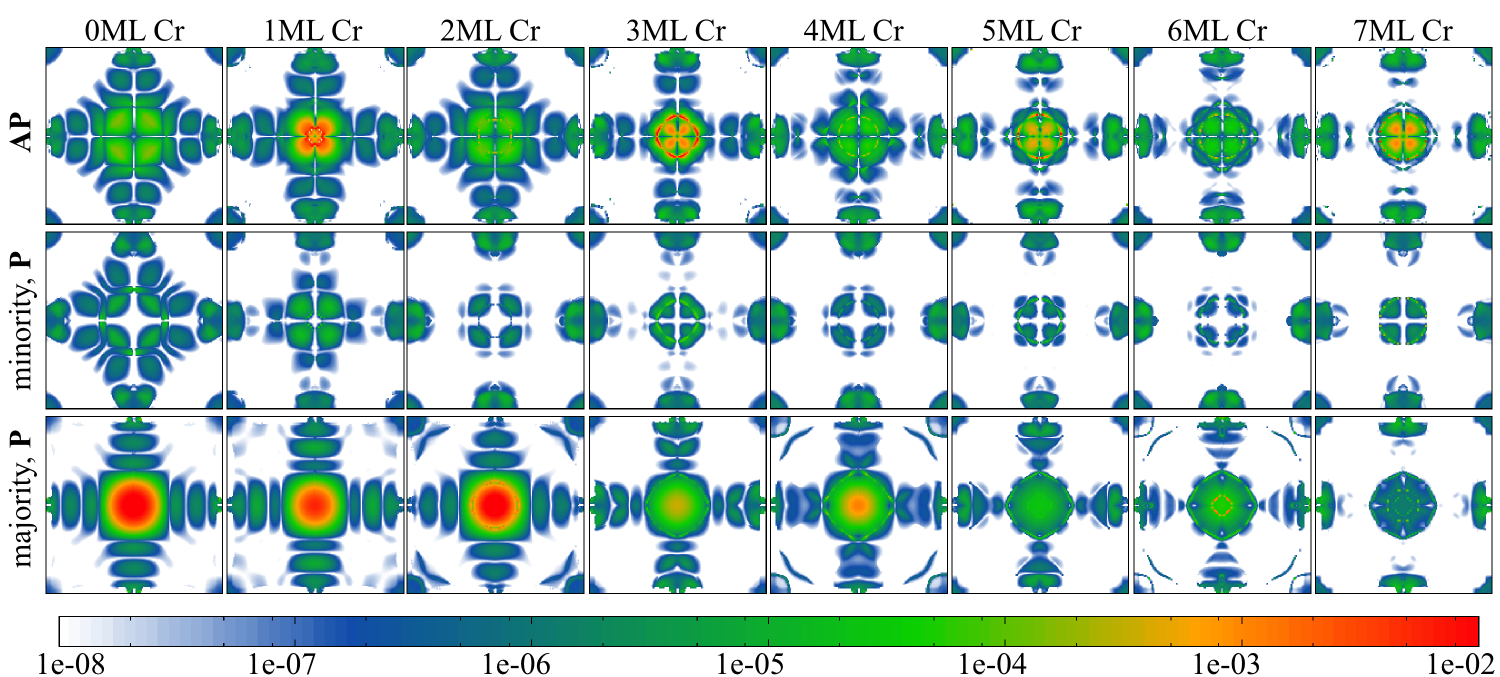}
  \caption{(Color online) Majority ($\mathrm{P}$, bottom row),
    minority ($\mathrm{P}$, middle row) and $\mathrm{AP}$ (top row)
    transmission $T(\kp, E_F)$ within two-dimensional Brillouin zones
    for Fe(001)/$x$(Cr)/6(MgO)/Fe(001) MTJs with Cr layer thicknesses
    of $x = 0, \ldots, \unit[7]{ML}$ (from left to right).  The
    two-dimensional Brillouin zones cover the range between $-\pi/a$
    and $\pi/a$.}
  \label{fig:fecrmgofe_maps}
\end{figure*}

There are two possible mechanisms that may explain other wavelengths than the $\unit[2]{ML}$.
Firstly, one could think of spin-density waves within the Cr
interlayers. Spin-density waves are found for Cr bulk
systems\cite{Fawcett1988} and are related to nesting vectors of the
Fermi surface.  Nesting vectors that come into question are shown in
Fig.~\ref{fig:fecrmgofe_fs} for a cross section of the Fermi surface
in the (100) plane. The corresponding wavelengths of these vectors
along the [001] direction (i.\,e.\ in transport and growth direction)
are given in Table~\ref{tab_nesting}.  The only vector that exhibits a
wavelength which is comparable to that of the transmission
($\unit[2.17]{ML}$) is S, with an oscillation period of
$\unit[2.12]{ML}$. However, since we are interested in the oscillatory
onset at $\kp=0$, the vector S cannot explain our findings because it is offset
from $\bar{\Gamma}$.

The oscillatory exponential decay of $T_{\mathrm{maj}}^{\mathrm{P}}(\kp=0)$ 
is explained most promisingly in terms of the complex
bandstructure\cite{Heine1965} of the Cr interlayers.  Since a
(continuous) dispersion relation is not defined for thin films, due to
lack of translational invariance, we refer to the complex
bandstructure of bulk Cr along [001].  The latter is decomposed with
respect to the irreducible representations of the point group $4mm$
($\Delta_1$, $\Delta_5$, $\Delta_2$ and $\Delta_{2'}$) of the
associated Bloch states (Fig.~\ref{fig:fecrmgofe_bs}).

A complex band structure of a periodic system is the conventional band
structure extended to Bloch vectors ($\kp, k_{\perp}$) with complex
wave numbers $k_{\perp}$. The associated bands can be cast into four
categories\cite{Chang1982}: (i) \textit{real bands} which correspond
to the conventional band structure and have $\mathrm{Im} k_{\perp} = 0$; (ii)
\textit{imaginary bands of the first kind} have $\mathrm{Re} k_{\perp} =
0$ and $\mathrm{Im} k_{\perp} \neq 0$; (iii) \textit{imaginary bands of the
  second kind} with $\mathrm{Re} k_{\perp} = \nicefrac{\pi}{a}$ and $\mathrm{Im}
k_{\perp}\neq 0$; and $(iv)$ \textit{complex bands} with $\mathrm{Re} k_{\perp}
\neq 0$, $\mathrm{Re} k_{\perp} \neq\nicefrac{\pi}{a}$ and $\mathrm{Im} k_{\perp} \neq
0$.

The imaginary part of $k_{\perp}$ is denoted as $\kappa$ and
represents a measure for the decay rate of evanescent Bloch 
states\footnote{Since $k_{\perp}$ is commonly defined in units of 
$\nicefrac{\pi}{a}$ and $a$ corresponds to an interlayer distance of 
$\unit[2]{ML}$, a factor of $\nicefrac{\pi}{2}$ has to be incorporated 
to obtain it in units of $\nicefrac{1}{\mathrm{ML}}$.}. At the Fermi energy, a
complex band of the second kind shows up at $k_{\perp} = (0.922,
0.503)\nicefrac{\pi}{a}$ (circles in Fig.~\ref{fig:fecrmgofe_bs}).  The
corresponding decay rate of $\kappa = 0.78 / \unit{ML}$ matches well
the estimated exponential decay of the majority transmission ($\kappa
= 0.75 / \unit{ML}$, Fig.~\ref{fig:cr_conductance_gamma}a).  Due to
the nonvanishing real part the exponential decay exhibits an
oscillatory envelope with a wavelength $\lambda =
\nicefrac{\pi}{\mathrm{Re}[k_{\perp}]} \approx \unit[2.17]{ML}$ which
agrees well with the fit in Fig.~\ref{fig:cr_conductance_gamma}b.  We
conclude therefore that the thickness dependence of the transmission
for the majority states at the $\bar{\Gamma}$ point is very likely
governed by this $\Delta_1$ state, provided the electronic structure
of ultra-thin Cr films is well described by that of bulk Cr.

\begin{figure}[hb]
  \includegraphics[width=0.74\columnwidth]{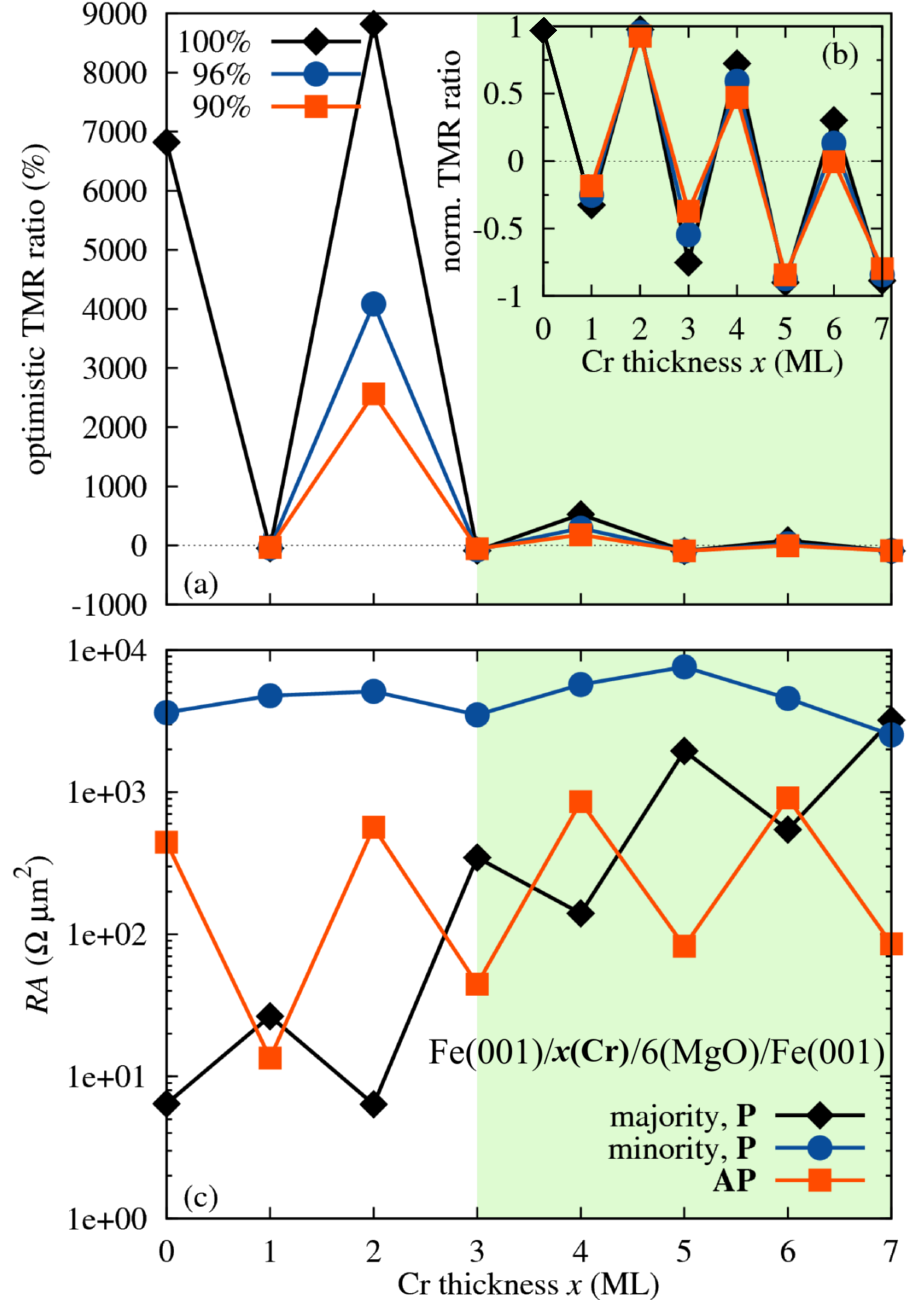}
  \caption{(Color online) Cr thickness dependence of (a) optimistic
    and (b) normalized TMR ratios in Fe(001)/$x$(Cr)/6(MgO)/Fe(001)
    MTJs. Since variations of the Cr layer thickness cannot be ruled
    out in experiment, a model with resistors in parallel connection
    is assumed to mimic Cr-thickness fluctuations (line styles
    indicate the weight $w$; see text).  (c) Resistance area product
    $RA$ for parallel magnetic ($\mathrm{P}$: majority, minority) and
    anti-parallel magnetic configurations ($\mathrm{AP}$), shown on a
    logarithmic scale.}
  \label{fig:fecrmgofe}
\end{figure}

\begin{figure*}[hbt]
  \includegraphics[width=0.8\textwidth]{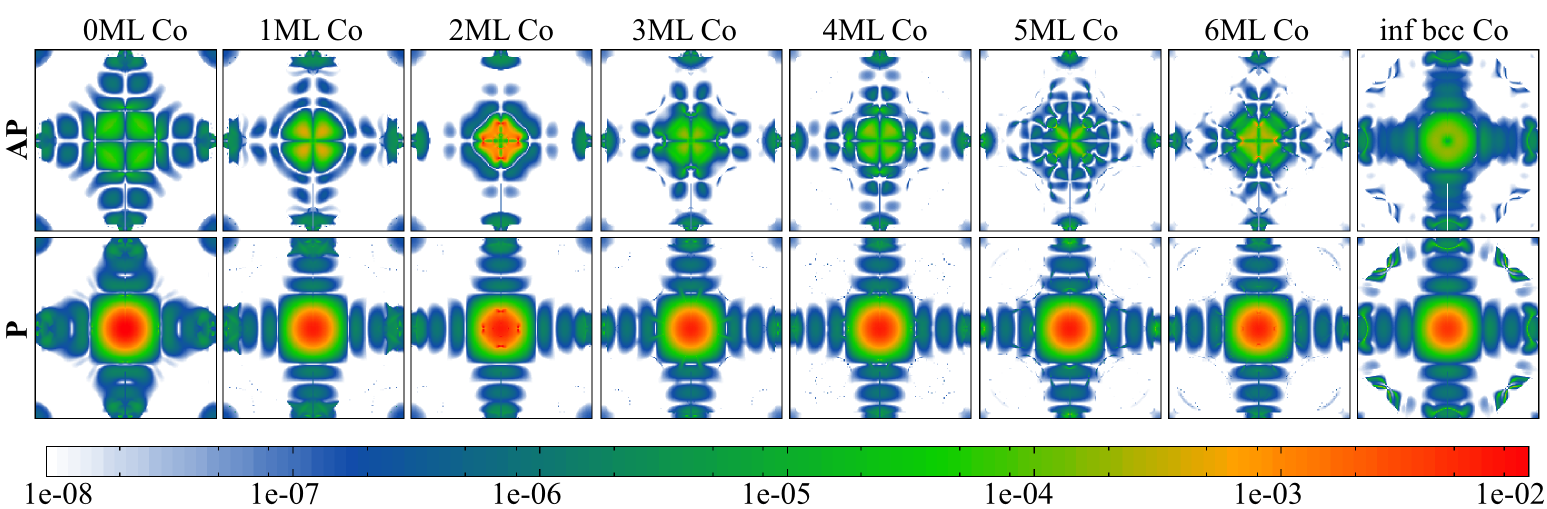}
  \caption{(Color online) Transmission $T(\kp, E_F)$ for the
    $\mathrm{P}$ (bottom row) and $\mathrm{AP}$ (top row) magnetic
    configuration within two-dimensional Brillouin zones for
    Fe(001)/$x$(Co)/6(MgO)/ $x$(Co)/Fe(001) MTJs, $x = 0, \ldots,
    \unit[6]{ML}$ (from left to the right). The panels on the right
    hand side are for Co(001)/6(MgO)/Co(001) MTJs with infinite bcc Co
    leads. The Brillouin zones cover the range between $-\pi/a$ and
    $\pi/a$.}
  \label{fig:fecomgocofe_maps}
\end{figure*} 

Transmission maps display the spin-resolved transmission for
$\mathrm{P}$ and $\mathrm{AP}$ versus $\kp$ for each Cr thickness $x$
(Fig.~\ref{fig:fecrmgofe_maps}).  As observed for $T^{\mathrm{P}}_{\mathrm{maj}}(\kp=0)$ one finds
within the majority transmission maps a clearly visible modulation of
$T^{\mathrm{P}}_{\mathrm{maj}}(\kp)$ for $\kp$-points in the center regions with maxima (minima)
at even (odd) multiples of $x$.  It is reasonable to assume that the
oscillations of these transmissions are caused by the same
bandstructure effect as it was discussed for $T^{\mathrm{P}}_{\mathrm{maj}}$ at
$\bar{\Gamma}$. Furthermore, the majority $RA$ products in
Fig.~\ref{fig:fecrmgofe} exhibit even-odd oscillations as well,
indicating constructive superposition of the oscillations of the
individual $T^{\mathrm{P}}_{\mathrm{maj}}(\kp)$.  

Although the cloverleaf-like structures within the minority
transmission maps in Fig.~\ref{fig:fecrmgofe_maps} exhibit slight
variations which are in anti-phase to the majority transmission
modulations, the corresponding $RA$ products in
Fig.~\ref{fig:fecrmgofe} reveal no signatures of such an even-odd
characteristic.  The weak thickness dependence can be understood with
help of the complex bandstructure in Fig.~\ref{fig:fecrmgofe_bs}.
From first-principles investigations on Fe/MgO/Fe it is
known\cite{Heiliger2006,Butler2008} that $\Delta_5$ states are the
main carrier within the minority transport channel. Due to the
$\Delta_5$ \textit{real band} at the Fermi energy it is very likely
that these Bloch states just propagate undamped through the Cr
interlayer. The minority $RA$ is therefore marginally affected by the
Cr spacer and gives contributions which are similar to those of 
Fe/MgO/Fe MTJs.

In contrast to the above finding, the pronounced modulations within
the $\mathrm{AP}$ transmission maps in Fig.~\ref{fig:fecrmgofe_maps}
--- with maxima (minima) at odd (even) $x$ --- lead to an even-odd
oscillation of the corresponding $RA$.  
Due to the spin-filter effect of the Cr interface layer $RA^{\mathrm{AP}}$
is in anti-phase to $RA_{\mathrm{maj}}^{\mathrm{P}}$. This behavior results
in TMR ratios which exhibit $\unit[2]{ML}$ oscillations 
with periodic changes of the sign. The amplitudes of the TMR ratios, are with about
$\unit[7000]{\%}$ and $\unit[9000]{\%}$ at $x = \unit[0]{ML}$ and
$\unit[2]{ML}$, considerably larger than for the other thicknesses, with
values between about $-\unit[100]{\%}$ and $+\unit[100]{\%}$. This
even-odd oscillation of the TMR ratio as a function of the Cr
thickness has been observed experimentally\cite{Matsumoto2009} but
with a phase shift of $\unit[1]{ML}$.  In more recent
experiments\footnote{R. Matsumoto and S. Yuasa, private
  communication.} it has been found that this phase shift depends on
whether the Cr interlayer is grown after or before growth of the MgO
barrier. However, a large maximum of the TMR ratio for small Cr
thicknesses does not show up in both growth conditions.  Instead, an
exponential decay of the optimistic TMR ratio for increasing $x$ is
reported\cite{Matsumoto2009}. Since variations of the Cr layer thickness cannot be ruled
out in experiment, we assume a model with resistors in parallel
connection to mimic Cr-thickness fluctuations.  The resistance of the
mean thickness $x$ is weighted by $w$, the contributions from $x-1$
and $x+1$ are weighted by $(\unit[100]{\%} - w) / 2$, respectively.
With already large central weights of $w = \unit[96]{\%}$ and $w =
\unit[90]{\%}$ this model is able to reproduce the principal
experimental TMR characteristics\cite{Matsumoto2009}.  For a detailed
analysis of the effect of structural imperfections, however, one has
to rely on more sophisticated computational approaches, like the
coherent potential approximation or a supercell method.


%

\subsection{Fe(001)/ x(Co)/ 6(MgO)/ x(Co)/ Fe(001)}
\label{subsec:Co}

In this section we discuss the effects of Co interlayers embedded into
Fe(001)/MgO/Fe(001) MTJs on the electronic transport. With respect to
previous theoretical investigations of MgO barriers with bcc Co
leads\cite{Zhang2004a} we specifically studied the thickness
dependence of ultrathin Co interlayers which are inserted at
\textit{both} Fe/MgO interfaces. Since Co grows epitaxially only up to
few ML on a bcc substrate, the Co thicknesses $d_{\mathrm{Co}}$ are
restricted to $x\leq\unit[6]{ML}$. The monolayer
separations are taken identical to the case of Cr interlayers.
Substituting all Fe atoms with Co atoms the effect of semi-infinite 
bcc Co leads is studied in addition.

First, we recall previously reported conductances
and TMR ratios in MTJs with equal thickness variations of both Co
interlayers \cite{Bose2009}. The corresponding $\kp$-resolved
transmissions versus $x$ are shown in Fig.~\ref{fig:fecomgocofe_maps}
for both magnetic configurations ($\mathrm{P}$, $\mathrm{AP}$). The
shapes of the transmission maps for $\mathrm{P}$ configuration (bottom
row) are --- beside the case of the semi-infinite Co leads --- very
similar. In contrast to this weak dependence, one observes a rise and
a decline of the transmission probabilities in the central regions of the Brillouin zones
for the $\mathrm{AP}$ configuration for small $x$ (top row). The maximum
shows up at $x=\unit[2]{ML}$. Based on these observations
it is reasonable to expect a relatively constant
behavior of the $\mathrm{P}$ conductance and a maximum at
$x=\unit[2]{ML}$ for the $\mathrm{AP}$ conductance.

\begin{figure}[tb]
  \includegraphics[width=0.69\columnwidth]{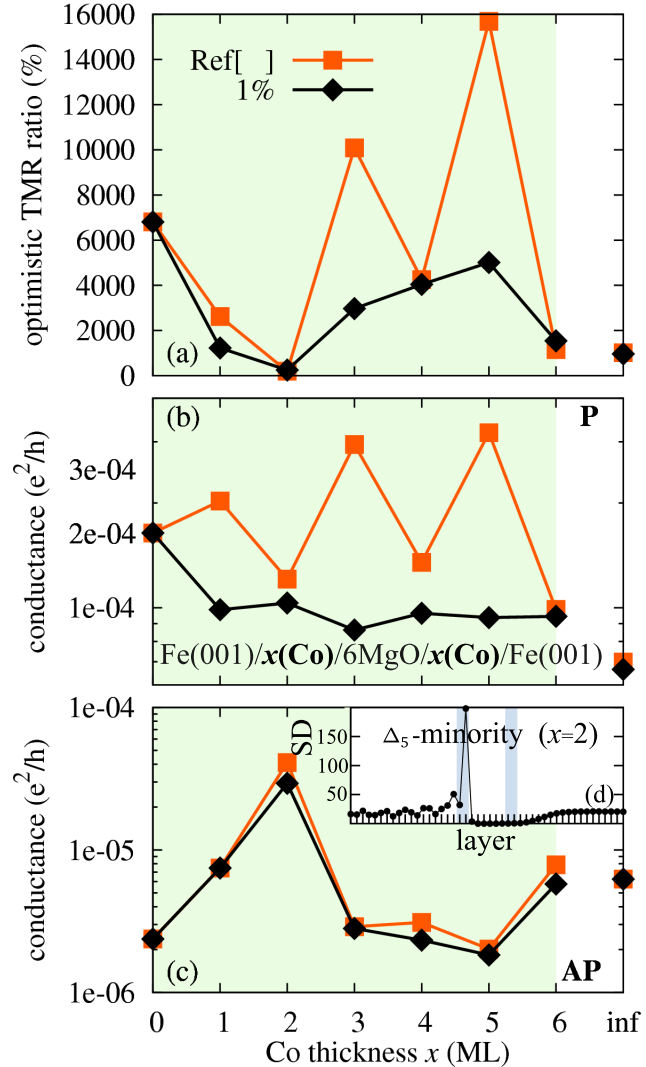}
  \caption{(Color) Co thickness dependence of (a) optimistic TMR
    ratio, $\mathrm{P}$ (b), and $\mathrm{AP}$ (c) conductances in
    symmetric Fe(001)/$x$(Co)/6(MgO)/$x$(Co)/Fe(001) MTJs.  Red
    squares, taken from Ref.~\onlinecite{Bose2009}, are for data which
    exhibit hotspots in the transmission probabilities $T(\kp,E_F)$. The
    black symbols show the
    data with these hot spots being removed (see text). Panel (d)
    displays the layer-resolved spectral density (in states/Hartree)
    of a $\Delta_5$-minority interface resonance with
    $\kp=(0.095,0.008)\nicefrac{\pi}{a}$. The blue areas mark the
    position of the $\unit[2]{ML}$ thick Co interlayers within the MTJ\@.}
  \label{fig:fecomgocofe}
\end{figure}

The elsewhere published conductances and TMR ratios\cite{Bose2009} are
inserted into Fig.~\ref{fig:fecomgocofe} as red boxes.  In
accordance with the findings for bcc Co leads\cite{Zhang2004a}, we
identify specific Co thicknesses ($x=\unit[3]{ML}$ and $\unit[5]{ML}$)
that exhibit larger TMR values than those obtained for pure Fe/MgO/Fe
junctions. In particular, the TMR ratio follows the even-odd type
characteristic which shows up for the conductance $C^{\mathrm{P}}$.
But, this behavior of $C^{\mathrm{P}}$ does not reflect the weak
thickness dependence as expected from the transmission probabilities
(Fig.~\ref{fig:fecomgocofe_maps}).  It turned out that this even-odd
change is considerably affected by single interface resonances whose
transmission probabilities contribute with up to $\unit[70]{\%}$ to
the conductance.  These hot spots within the transmission maps occur
preferably in ideal, symmetric MTJs at zero bias
voltage\cite{Belashchenko2005}. They are strongly diminished by
breaking the symmetry, for instance by means of a tiny bias voltage
or structural imperfections of the sample.
A bias voltage of $\unit[0.02]{V}$ is sufficient to
destroy the resonant states and to reduce the corresponding
transmission probabilities by several orders of magnitude.

The total number of hot spots in each transmission map of
Fig.~\ref{fig:fecomgocofe_maps} is less than $10$. Instead of removing
the resonances by a small bias voltage, the transmissions of these
states are identified and neglected.  These filtered data are shown as
black diamonds in Fig.~\ref{fig:fecomgocofe}. The effect of the
resonances shows up mainly for $C^{\mathrm{P}}$.  For the latter one
obtains, after an initial decrease of about $\unit[50]{\%}$ from
$x=\unit[0]{ML}$ to $\unit[1]{ML}$, the expected nearly constant behavior.

The filtered $\mathrm{AP}$ conductances agree well with unfiltered
ones, in particular the maximum at $x=\unit[2]{ML}$. Inspecting the
corresponding transmission map (Fig.~\ref{fig:fecomgocofe_maps}), this
maximum is attributed to enhanced transmissions in the center region
of the 2BZ which are caused by $\Delta_5$-minority interface resonances
within the Co interface layer (Fig.~\ref{fig:fecomgocofe}d). In
contrast to the transmission resonances (hot spots), these transmissions
are unaffected by small bias voltages.

Since the thickness dependence of $C^{\mathrm{P}}$ is weakened by the
filter procedure, that of the TMR ratio is as well. In particular, the
maxima at $x=\unit[3]{ML}$ and $\unit[5]{ML}$ do not show up.  The
minimum of approximately $200\%$ at $x=\unit[2]{ML}$ is, however, still
present; it is explained by the increase of $C^{\mathrm{AP}}$ --- up
to a value comparable with $C^{\mathrm{P}}$ --- due to the
$\Delta_5$-minority interface resonances. For increasing $x$ the TMR ratio
increases monotonously up to a maximum at $x=\unit[5]{ML}$, with
$\unit[4600]{\%}$ considerably smaller than the $\unit[7000]{\%}$
obtained for pure Fe/MgO/Fe MTJs.
\begin{figure}[b]
  \includegraphics[width=0.69\columnwidth]{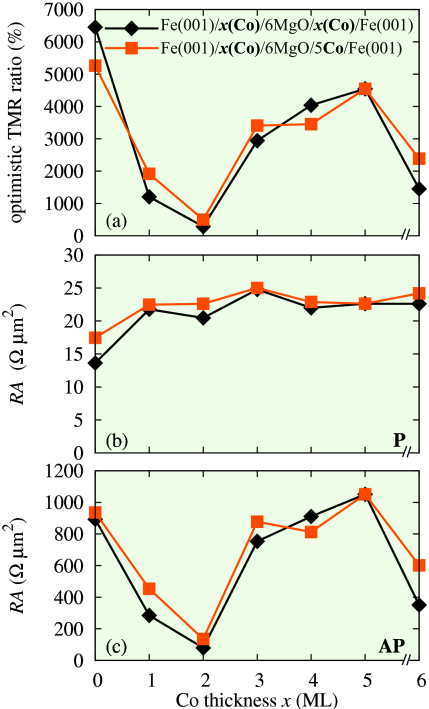}
  \caption{(Color) Optimistic TMR ratios (a) and resistance-area
    products $RA$ for $\mathrm{P}$ (b) and $\mathrm{AP}$ (c) of
    symmetric (black) and asymmetric (red)
    Fe(001)/$x$(Co)/6(MgO)/$y$(Co)/Fe(001)MTJs. Hot spots were
    filtered out (see e.\,g.\ Fig.~\ref{fig:fecomgocofe}).}
  \label{fig:fe5comgoXcofe}
\end{figure}

We note in passing that the enhanced TMR ratio in MTJs with
semi-infinite bcc Co leads, reported in Ref.~\onlinecite{Zhang2004a},
are not reproduced.  Previous investigations of Fe/MgO/Fe systems have
shown that conductances depend strongly on details of the
calculations, in particular on the atomic positions in the interface region.
Thus, one is lead to attribute the above mentioned discrepancy to
computational details.

Since increased TMR ratios are not obtained by symmetric MTJs, we
studied the effect of asymmetric MTJs, with one interlayer thickness
fixed to $\unit[5]{ML}$, in addition.  The fixed thickness of
$\unit[5]{ML}$ is chosen with respect to the largest TMR ratio in
symmetric MTJs (Fig.~\ref{fig:fecomgocofe}).  Both the $RA$ products
and the TMR ratios of the asymmetric junctions do not differ
significantly from those of the symmetric junctions
(Fig.~\ref{fig:fe5comgoXcofe}). For both kinds of MTJs, the TMR ratio
follows the characteristics of $RA^{\mathrm{AP}}$ because
$RA^{\mathrm{P}}$ is almost constant. The presence of the minimum at
$x=\unit[2]{ML}$ additionally substantiates that the
$\Delta_5$-minority interface resonances, which cause the drop of
$RA^{\mathrm{AP}}$, are marginally affected by symmetry breaking of an
ideal MTJ\@.

\section{Conclusion}
In summary, the computed conductances and TMR ratios exhibit the oscillatory 
decays for thickness variations of ultrathin Cr interlayers, 
which have been found in experiment. The analysis of the associated
transmission probabilities reveals that the tunneling of 
Bloch states is affected by the interplay of two mechanisms. On the one hand
a spin-filter effect which is induced by the enhanced magnetic moments
of the Cr interface layers and on the other hand the presence of
complex bands which are formed within the Cr interlayers.
The oscillations are therefore mixtures
of $\unit[2]{ML}$ oscillations of magnetic origin and superpositions of the individual
modulations of the tunneling Bloch states, which can be traced back to
the corresponding complex wave vectors. Our results further indicate that
spin-density waves are of minor importance for understanding of electronic
transport through Fe/MgO/Fe MTJs with ultrathin Cr interlayers. 

The embedding of Co interlayers at both interfaces does not lead to
an increase of the TMR ratios with respect to Fe/MgO/Fe MTJs. Please note that
the reference values for $x=0$ (i.e. no Co interlayer) have been calculated under the assumption of
ideal interface structures and are therefore probably considerably
overestimated. Thus, we suggest to include 
the effects of imperfect interfaces, which are unavoidable in real
samples, in a future investigation.

\begin{acknowledgments}
  Helpful discussions with S. Yuasa, R. Matsumoto, and N.\,F. Hinsche
  are gratefully acknowledged. This work is supported by the
  Sonderforschungsbereich 762, Functionality of Oxidic Interfaces.
  P.\,B. is supported by the International Max Planck Research School
  for Science and Technology of Nanostructures.
\end{acknowledgments}

\bibliographystyle{apsrev_without_url}
\bibliography{literature}

\end{document}